# Magnetic Monopole Interactions, Chiral Symmetries, and the NuTeV Anomaly

## Jay R. Yablon[*]

910 Northumberland Drive

Schenectady, New York, 12309-2814

**The long-sought "magnetic monopole" appears to be – not a fermion distinct from the electrons and the quarks – but** *a charge carried by the known electrons and the quarks themselves*. **Similarly to the "Z" charge of electroweak theory, however, this magnetic monopole charge only manifests its interactions at sufficiently-high energy, and its interactions may not respect chiral symmetry. Calculated cross-section enhancements from magnetic monopole interactions at s=M$_z$$^2$ reduce the observed weak mixing angle for $e\bar{e} \to \mu\bar{\mu}$ decays by about $\Delta \sin^2 \theta_W(M_Z) = -.0030$ relative to $\nu\bar{\nu} \to \mu\bar{\mu}$ decays, and so may help account for the NuTeV anomaly. Similar, though less-pronounced reductions are calculated for $e\bar{e} \to q\bar{q}$.**


---


[*] jyablon@nycap.rr.com




# 1. Introduction

In an earlier paper, [1] the author demonstrated how to break the symmetry of a duality-invariant Lagrangian so that at low energy, electric monopole interactions continue to be observed but magnetic monopole interactions become very highly suppressed to the point of effectively vanishing. In the course of this development, it was found that local duality symmetry combined with local U(1)$_{EM}$ gauge symmetry leads naturally and surprisingly to an SU(2)$_D$ duality gauge group.

In this paper, which continues the development of [1] with a calculation of partial and full widths and cross sections for magnetic monopole interactions, we take a close look at the magnetic monopoles themselves as particles which presumably have a spin ½, fermion character, in an effort to directly answer the long-standing question: what, exactly, is a magnetic monopole? The answer yields some surprising insights into the nature of chiral symmetries and asymmetries, and may point toward a better understanding of the so-called NuTeV anomaly.

# 2. A Brief Review of Duality Symmetry Breaking in Maxwell's Electrodynamics

In [1], the author demonstrated how to break the duality symmetry of the duality-symmetric Lagrangian (see [1], equation (5.1)):[*]

$$L = \overline{\psi}_e(i\gamma^\mu\partial_\mu - m_e)\psi_e + \overline{\psi}_m(i\gamma^\mu\partial_\mu - m_m)\psi_m - g_e J^\mu A_\mu - g_m P^\mu M_\mu - \tfrac{1}{4}F_{\mu\nu}F^{\mu\nu} - \tfrac{1}{4}{}^*F_{\mu\nu}{}^*F^{\mu\nu}, \quad (2.1)$$

by rotating the interaction term $-g_e J^\mu A_\mu - g_m P^\mu M_\mu$ through complexion angle $\alpha$ via (5.8):

$$L_{gJB} = -g_e J^\mu A_\mu - g_m P^\mu M_\mu = -\begin{pmatrix} A_\mu & M_\mu \end{pmatrix}\begin{pmatrix} g_e J^\mu \\ g_m P^\mu \end{pmatrix}$$

$$\rightarrow L_{gJB}' = -g_e' J^{\mu'} A_\mu' - g_m' P^{\mu'} M_\mu' = -\begin{pmatrix} A_\mu' & M_\mu' \end{pmatrix}\begin{pmatrix} g_e' J^{\mu'} \\ g_m' P^{\mu'} \end{pmatrix}. \qquad (2.2)$$

where, from (5.7)):

$$\begin{pmatrix} g_e J^\nu \\ g_m P^\nu \end{pmatrix} \rightarrow \begin{pmatrix} g_e' J^{\nu'} \\ g_m' P^{\nu'} \end{pmatrix} = \begin{pmatrix} \cos\alpha & \sin\alpha \\ -\sin\alpha & \cos\alpha \end{pmatrix}\begin{pmatrix} g_e J^\nu \\ g_m P^\nu \end{pmatrix}, \qquad (2.3)$$

and from (5.9):

---

[*] In this section, unless otherwise indicated, references are to equation numbers in [1]. Notations throughout are the same as those employed in [1].



$$\begin{pmatrix} A_\mu \\ M_\mu \end{pmatrix} \rightarrow \begin{pmatrix} A_\mu' \\ M_\mu' \end{pmatrix} = \begin{pmatrix} \cos\alpha & \sin\alpha \\ -\sin\alpha & \cos\alpha \end{pmatrix} \begin{pmatrix} A_\mu \\ M_\mu \end{pmatrix}, \tag{2.4}$$

describe the transformation properties of the duality doublets $\begin{pmatrix} g_e J^\nu \\ g_m P^\nu \end{pmatrix}$ and $\begin{pmatrix} A_\mu \\ M_\mu \end{pmatrix}$. We shall refer to all of these as the "unmixed" charges $g_e$, $g_m$, currents $J^\nu$, $P^\nu$, and "bosons" $A_\mu$, $M_\mu$, to be later distinguished from "observable" charges, currents, and bosons that these are rotated into via a duality transformation followed by duality symmetry breaking.

Then, by *imposing* (not deriving) an electroweak-like *symmetry-breaking condition* (5.11) we established the *observed* electric current:

$$J_{em}{}^\mu \equiv J^{\mu\prime} = J^\mu + P^\mu, \tag{2.5}$$

we established an *observable* magnetic monopole current (5.14):

$$P^{\mu\prime} = P^\mu - \sin^2\alpha \cdot J^{\mu\prime} \tag{2.6}$$

and we established the respective running charge relationships (5.12) and (5.15):

$$e = g_e' = \cos\alpha \cdot g_e = \sin\alpha \cdot g_m, \tag{2.7}$$

$$g_m' = \frac{g_m}{\cos\alpha} = \frac{g_e}{g_e'} g_m, \tag{2.8}$$

where the *unit* charges are related by Dirac's quantization condition (5.14) for n=1:

$$g_e' \cdot g_m' = g_e \cdot g_m = 2\pi \hbar c. \tag{2.9}$$

The *observed* running charges are the "primed" charges $g_e' \equiv e$ and $g_m'$. This is all very similar to how the observed electroweak neutral current is established, see discussion following (5.9). It was pointed out in [1], that it is the imposition of (2.5) which ensures that (2.9) holds. In other words, the electroweak-like condition (2.5) appears to be required to ensure that the Dirac quantization condition emerges invariantly from this rotation and symmetry breaking.

In the above, $J^\mu = \bar{\psi}_e \gamma^\mu Q_e \psi_e$ is the unmixed current four vector for the "electric monopole" Dirac spinor wavefunction $\psi_e$, while $P^\mu = \bar{\psi}_m \gamma^\mu Q_m \psi_m$ is the unmixed current four vector for the "magnetic monopole" Dirac spinor wavefunction $\psi_m$ (see discussion following (5.6)), *before* these are mixed together to form observable currents according to (2.5) and (2.6) above. $Q_e$ is taken to be an "unmixed" electric charge generator, and $Q_m$ is taken to be an "unmixed" magnetic charge generator. Now, we take a closer look at the mixing of currents as set forth in (2.5) and (2.6) above.



## 3. Are Magnetic Charges Carried By Distinct Fermions, or do Electrically-Charged Fermions Also Carry Magnetic Charges?

We begin the present development by exploring the *observed* electric current of (2.5), which we write as:

$$J_{em}{}^\mu = \overline{\psi}\gamma^\mu Q\psi \equiv J^{\mu\prime} = J^\mu + P^\mu = \overline{\psi}_e\gamma^\mu Q_e\psi_e + \overline{\psi}_m\gamma^\mu Q_m\psi_m, \tag{3.1}$$

as well as the *observable* magnetic current (2.6) which we write as:

$$P^{\mu\prime} = P^\mu - \sin^2\alpha \cdot J_{em}{}^\mu = \overline{\psi}_m\gamma^\mu Q_m\psi_m - \sin^2\alpha \cdot \overline{\psi}\gamma^\mu Q\psi, \tag{3.2}$$

where Q is the *observed* electric charge generator. Trying to interpret equation (3.1) presents an immediate dilemma. Because we have mixed the electric current $J^\mu$ with a magnetic current $P^\mu$ to construct the *observed* electric current $J_{em}{}^\mu = J^{\prime\mu} = J^\mu + P^\mu$ (as well as to construct an observable magnetic current $P^{\mu\prime}$ in (3.2)), we know that it is the $\psi, \overline{\psi}$ in $\overline{\psi}\gamma^\mu Q\psi$ in (3.1), stripped of any "e" or "m" label, which must represent the *observed* fermions. And, we know that it is the electric charge generator Q in $\overline{\psi}\gamma^\mu Q\psi$, stripped of any "e" or "m" label, which must represent the observed electric charge generator given for three generations of electron by $\psi_{electron} \equiv |Q = -1\rangle$, up quark by $\psi_u \equiv |Q = \frac{2}{3}\rangle$, and down quark by $\psi_d \equiv |Q = -\frac{1}{3}\rangle$.

Yet, $\overline{\psi}_e\gamma^\mu Q_e\psi_e + \overline{\psi}_m\gamma^\mu Q_m\psi_m = \overline{\psi}\gamma^\mu Q\psi$ in (3.1) contains what appear to be *separate and distinct fermions* $\overline{\psi}_e, \psi_e$ and $\overline{\psi}_m, \psi_m$ which combine in some fashion to form the observed fermions $\psi, \overline{\psi}$. Additionally, these terms contain separate electric and magnetic charge gauge generators $Q_e$ and $Q_m$ combining to form the observed electric charge generator Q for the observed Fermions. Of course, this was part of the initial hypothesis introduced by the author in equation (2.1) and (3.5) of [1]. Now, we must take hard look and figure out how to interpret this.

To gain our bearings, we contrast the above to the analogous neutral currents in electroweak theory, including their chiral vertex factors. For (3.1), we contrast:

$$\begin{aligned} J_{em}{}^\mu &= \overline{\psi}\gamma^\mu Q\psi = J_Y{}^\mu + J_3{}^\mu = \overline{\psi}\gamma^\mu \tfrac{1}{2}(c_V(Y) - c_A(Y)\gamma^5)\psi + \overline{\psi}\gamma^\mu \left(\tfrac{1}{2}(1-\gamma^5)I_3\right)\psi \\ &= \overline{\psi}\gamma^\mu \tfrac{1}{2}(c_V(Y) - c_A(Y)\gamma^5) + \gamma^\mu \left(\tfrac{1}{2}(1-\gamma^5)I_3\right)\psi \end{aligned} \tag{3.3}$$

and for (3.2) we contrast:

$$\begin{aligned} J_Z^\mu &= J_3^\mu - \sin^2\theta_W \cdot J_{em}^\mu = \overline{\psi}\gamma^\mu \tfrac{1}{2}(1-\gamma^5)I_3\psi - \sin^2\theta_W \cdot \overline{\psi}\gamma^\mu Q\psi \\ &= \overline{\psi}\gamma^\mu \left(\tfrac{1}{2}(1-\gamma^5)I_3 - \sin^2\theta_W \cdot Q\right)\psi \equiv \overline{\psi}\gamma^\mu \tfrac{1}{2}(c_V(Z) - c_A(Z)\gamma^5)\psi \end{aligned} \tag{3.4}$$

The terms $\overline{\psi}\gamma^\mu Q\psi$ without $\gamma^5$ in both of the above tell us that the *observed* electromagnetic current is chiral symmetric, while $\overline{\psi}\gamma^\mu \tfrac{1}{2}(1-\gamma^5)I^3\psi \equiv \overline{\psi}\gamma^\mu \tfrac{1}{2}(c_V(I^3) - c_A(I^3)\gamma^5)\psi$, hence



$c_V(I^3) = c_A(I^3) = I^3$, tell us that the weak isospin currents are V-A currents which maximally violate parity.

But for the moment, it is most important to note while the first line in each of (3.3) and (3.4) is very similar to equations (3.1) and (3.2), we are able in the second line in each of (3.3) and (3.4) to combine terms in a way that does not appear possible for (3.1) and (3.2). That is, we can combine terms in (3.3) and (3.4) because the $\psi, \bar{\psi}$ are the same fermions for all of $J_{em}^{\mu}$, $J_Z^{\mu}$, $J_Y^{\mu}$ and $J_3^{\mu}$. Each fermion in any given chiral state (V, A, R, L) has an associated generator for electromagnetic charge $Q = Y + I_3$, "Z-charge" $Z = I_3 - \sin^2\theta_W \cdot Q$, weak hypercharge Y, and weak isospin charge $I_3$. In $J_Y^{\mu} = \bar{\psi}\gamma^{\mu}\frac{1}{2}(c_V(Y) - c_A(Y)\gamma^5)\psi$, we regard $\psi, \bar{\psi}$ as the same fermions in $J_3^{\mu} = \bar{\psi}\gamma^{\mu}(\frac{1}{2}(1-\gamma^5)T^3)\psi$ or in $J_{em}^{\mu} = \bar{\psi}\gamma^{\mu}Q_e\psi$ or in $J_Z^{\mu\prime} = \bar{\psi}\gamma^{\mu}(\frac{1}{2}(1-\gamma^5)I_3 - \sin^2\theta_W \cdot Q_e)\psi$. Equations (3.3) and (3.4), second line, make clear how these various charges are combined, *as well as the obvious but very important fact that a single Fermion carries several different types of charge*. (We have not even mentioned color charge, but of course, the quarks carry that as well, while the leptons carry a lepton number.)

Equations (3.1) and (3.2), in contrast, seem to be saying something different. Equation (3.1) seems to suggest that a Fermion wavefunction $\psi_e$ with an electric charge $Q_e$ is entirely different from a Fermion wavefunction $\psi_m$ with an electric charge $Q_m$. That is, equation (3.1) as written suggests that a magnetic monopole is a separate fermion (or are separate fermions) *distinct from the known fermions*, as opposed to yet another charge carried by the known fermions in addition to the Z and the Y and the $I^3$. But to better understand (3.1), we must now call this assumption into question.

In particular, if we wanted to add a second line to equations (3.1) and (3.2) analogous to the second line in (3.3) and (3.4), then we would have drop the distinction between $\psi_e$ and $\psi_m$ and no longer regard these as two separate fermions, one with an electric charge and the other with a magnetic charge. Instead, we would have to regard the $\psi_e$ and $\psi_m$ as a single Fermion *which has both an electric charge $Q_e$ and a magnetic charge $Q_m$*, just as we know that each fermion has an electromagnetic charge Q *and* a Z-charge Z *and* a weak hypercharge Y *and* a weak isospin charge $I_3$ *and* a color R, Y, B or lepton number L. This raises the question:

Is it possible that the *observable* magnetic monopoles, rather than being distinct fermions, are in fact charges carried by all of the same fermions – electrons and quarks – which carry electric charges (electron and quarks)? That is, might it be that the electron and the quarks *also carry a magnetic charge* that we simply have not yet observed because we have not yet reached sufficiently-high experimental energies for this magnetic charge to interact with the mediating 2.35 TeV vector boson predicted in [1], see just after (8.14)? And is it possible that this might be the answer to the long-standing question, what are magnetic monopoles and how might they manifest themselves to our experimental observation? We turn now to this very central question.

## 4. A Possible Connection Between Magnetic Monopoles and Chiral Symmetry

To explore this question, we now *hypothesize* that $\psi_e$ and $\psi_m$ are in fact not distinct fermion wavefunctions but rather are one and the same wavefunction, which we designate



simply as $\psi$. We then explore what the consequence might be of such a hypothesis. If we express this hypothesis mathematically by setting $\psi = \psi_e = \psi_m$ in (3.1) and (3.2), this would enable us to add a "second line" to (3.1) and (3.2) to more closely match (2.3) and (2.4), that is:

$$J_{em}^{\mu} = \overline{\psi}\gamma^{\mu}Q\psi \equiv J^{\mu\prime} = J^{\mu} + P^{\mu} = \overline{\psi}\gamma^{\mu}Q_e\psi + \overline{\psi}\gamma^{\mu}Q_m\psi ,$$
$$= \overline{\psi}\gamma^{\mu}(Q_e + Q_m)\psi \quad (4.1)$$

and:

$$P^{\mu\prime} = P^{\mu} - \sin^2\alpha \cdot J_{em}^{\mu} = \overline{\psi}\gamma^{\mu}Q_m\psi - \sin^2\alpha \cdot \overline{\psi}\gamma^{\mu}Q\psi$$
$$= \overline{\psi}\gamma^{\mu}(Q_m - Q\cdot\sin^2\alpha)\psi \quad (4.2)$$

This is a little bit closer to (3.3) and (3.4), but it raises some new questions. In particular, (4.1) seems to suggest that $Q = Q_e + Q_m$. Thus, for example, given that $Q = -1$ for the electron, this might mean, should $Q_e$ and $Q_m$ happen to be the *same magnitude* for each fermion, that $Q_e = Q_m = -\frac{1}{2}$ for the electrons, and $Q_e = Q_m = \frac{1}{3}$ for the up quarks, and $Q_e = Q_m = -\frac{1}{6}$ for the down quarks. But, contrasting to (2.3) and (2.4), this still leaves something out. Particularly, (4.1) makes the *assumption* that the gauge currents $J^{\mu}$ and $P^{\mu}$ are *each chiral symmetric, individually*. In fact – just as in the electroweak parallel (3.3) – such an assumption is not necessary and may well be incorrect. All that really matters is that the combination $J_{em}^{\mu} = J^{\mu} + P^{\mu}$ be chiral symmetric, because that is the *observable* electromagnetic current. The $J^{\mu}$ and $P^{\mu}$ individually could very well *not* be chiral symmetric, and for complete generality, we ought to account for this possibility.

Let us therefore *not assume* that $J^{\mu}$ and $P^{\mu}$ are each chiral symmetric, and so recast (4.1) in the most general chiral form:

$$J_{em}^{\mu} = \overline{\psi}\gamma^{\mu}Q\psi \equiv J^{\mu\prime} = J^{\mu} + P^{\mu} = \overline{\psi}\gamma^{\mu}\tfrac{1}{2}(c_V(Q_e) - c_A(Q_e)\gamma^5)\psi + \overline{\psi}\gamma^{\mu}\tfrac{1}{2}(c_V(Q_m) - c_A(Q_m)\gamma^5)\psi$$
$$= \overline{\psi}\gamma^{\mu}(\tfrac{1}{2}(c_V(Q_e) + c_V(Q_m)) - \tfrac{1}{2}(c_A(Q_e) + c_A(Q_m))\gamma^5)\psi \quad (4.3)$$

We must, of course, constrain the *observed* electromagnetic current $J_{em}^{\mu}$ to be chiral symmetric, so (4.3) then yields the two general constraints:

$$Q = \tfrac{1}{2}(c_V(Q_e) + c_V(Q_m)). \quad (4.4)$$

$$c_A(Q_e) = -c_A(Q_m). \quad (4.5)$$

Individual chiral symmetry for $J^{\mu}$ and $P^{\mu}$ would require us to further set $c_A(Q_e) = c_A(Q_m) = 0$.

Let us now make a second hypothesis: that *for any given fermion, the electric charge generator is equal to the magnetic charge generator*. This would allow us to generate both electric and magnetic currents from the same gauge group, so that the electron would have an electric and magnetic charge of -1, the up quark would have +2/3 for each charge, and the down



quark would carry -1/3 for each charge. This not only simplifies the generating charges with a suitable gauge group, but it also is consistent with the Dirac quantization condition for a unit charge which we write as $Q_e e \cdot Q_m m = n \cdot 2\pi \hbar c$.* This second hypothesis is imposed by setting:

$$c_V(Q_e) = c_V(Q_m). \qquad (4.6)$$

Combined with (4.4), this yields:

$$Q = c_V(Q_e) = c_V(Q_m). \qquad (4.7)$$

So, for the electron, the electric and magnetic charge generators are specified by $Q = c_V(Q_e) = c_V(Q_m) = -1$. For the up quarks, $Q = c_V(Q_e) = c_V(Q_m) = \frac{2}{3}$. For the down quarks, $Q = c_V(Q_e) = c_V(Q_m) = -\frac{1}{3}$. For the neutrino, $Q = c_V(Q_e) = c_V(Q_m) = 0$. Thus, substituting (4.7) and (4.5) into (4.3) now yields:

$$J_{em}^{\mu} = \overline{\psi}\gamma^{\mu}Q\psi \equiv J^{\mu\prime} = J^{\mu} + P^{\mu} = \overline{\psi}\gamma^{\mu}\tfrac{1}{2}(Q + c_A(Q_m)\gamma^5)\psi + \overline{\psi}\gamma^{\mu}\tfrac{1}{2}(Q - c_A(Q_m)\gamma^5)\psi. \qquad (4.8)$$

Now, we need to think closely about the chiral structure of $J^{\mu}$ and $P^{\mu}$. Referring to (4.5), one choice, of course, is $c_A(Q_e) = c_A(Q_m) = 0$, chiral symmetry, in which case $J^{\mu} = P^{\mu} = \overline{\psi}\gamma^{\mu}\tfrac{1}{2}Q\psi$. But let us instead take the opposite tack. Let us now make a third hypothesis, that $c_A(Q_e) = -c_A(Q_m) \neq 0$, and in particular, that $|c_A| = |c_V|$, which in effect, means that $J^{\mu}$ and $P^{\mu}$ are to violate parity maximally. One of these currents, according to this hypothesis, must have a V-A structure, and if that is so, then according to (4.5), this means that the other current must be V+A.

Now, there is nothing that tells us for sure whether $J^{\mu}$ ought to be V+A and $P^{\mu}$ ought to be V-A, or vice versa. So we shall assume for the moment that either choice is possible, and will ultimately rely on experimental observation to tell us which way nature has chosen. Here, noting that $P^{\mu}$ in (3.2) is situated analogously to the V-A weak isospin current $J_3^{\mu}$ in (3.4), and given the close analogy that has been developed with electroweak theory, we shall follow completely through with the electroweak analogy and hypothesize that $P^{\mu}$ also has a V-A structure. But we shall also leave open the possibility that nature has made the opposite choice, and again, will reply on experiment to decide. Our task for now is to build upon these hypotheses sufficiently to establish a point of contact with experimental cross sections.

Based on the above, this means that (4.7) is now updated to read:

$$Q = c_V(Q_e) = c_V(Q_m) = -c_A(Q_e) = c_A(Q_m), \qquad (4.9)$$

and that (4.8) now reduces to:

---

* This condition really applies just to the electron. For the fractionally-charged quarks, the analogous condition is $Q_e e \cdot Q_m m = \tfrac{1}{9} n \cdot 2\pi \hbar c$, which effectively makes 1/3 the magnitude of the "unit" charge.



$$J_{em}{}^\mu = \bar{\psi}\gamma^\mu Q\psi \equiv J^{\mu\prime} = J^\mu + P^\mu = \bar{\psi}\gamma^\mu \tfrac{1}{2}(1+\gamma^5)Q\psi + \bar{\psi}\gamma^\mu \tfrac{1}{2}(1-\gamma^5)Q\psi$$
$$= \bar{\psi}_R \gamma^\mu Q\psi_R + \bar{\psi}_L \gamma^\mu Q\psi_L \tag{4.10}$$

Additionally, this means that:

$$J^\mu = \bar{\psi}\gamma^\mu \tfrac{1}{2}(1+\gamma^5)Q\psi = \bar{\psi}_R \gamma^\mu Q\psi_R \tag{4.11}$$

$$P^\mu = \bar{\psi}\gamma^\mu \tfrac{1}{2}(1-\gamma^5)Q\psi = \bar{\psi}_L \gamma^\mu Q\psi_L, \tag{4.12}$$

so that the *observable* magnetic monopole current now becomes:

$$P^{\mu\prime} = P^\mu - \sin^2\alpha \cdot J_{em}{}^\mu = \bar{\psi}_L \gamma^\mu Q\psi_L - \sin^2\alpha \cdot \bar{\psi}\gamma^\mu Q\psi$$
$$= \cos^2\alpha \cdot \bar{\psi}_L \gamma^\mu Q\psi_L - \sin^2\alpha \cdot \bar{\psi}_R \gamma^\mu Q\psi_R \tag{4.13}$$

Now we must ask: does all of this make sense? Does it make sense that a "unmixed" electric current $J^\mu$ might turn out to be identified with V+A right-handedness and an "unmixed" magnetic current might turn out to have a V-A left-handedness? Or vice versa? Or, if, from (4.8), we write out more generally:

$$J^\mu = \bar{\psi}\gamma^\mu \tfrac{1}{2}(Q + c_A(Q_m)\gamma^5)\psi, \tag{4.14}$$

$$P^\mu = \bar{\psi}\gamma^\mu \tfrac{1}{2}(Q - c_A(Q_m)\gamma^5)\psi. \tag{4.15}$$

does it make sense that the difference between an electric and a magnetic monopole might in some way be related merely to a difference in chiral structure? If so, then chiral symmetry becomes very closely related to the electric / magnetic symmetry, and chiral symmetry breaking becomes closely related to duality symmetry breaking. Let us explore further.

As was pointed out in [1] following (3.14), the current four vectors for $J^\nu = F^{\mu\nu}{}_{;\mu}$ and $P^\nu \equiv *F^{\mu\nu}{}_{;\mu}$ are not independent, but are connected through the duality relation $*F^{\sigma\tau} \equiv \tfrac{1}{2!}\varepsilon^{\delta\gamma\sigma\tau}F_{\delta\gamma}$ between their fields. Similarly, both $A_\nu$ and $M_\nu$ are related to the same field tensor $F_{\mu\nu}$, via $F_{\mu\nu} = A_{\nu;\mu} - A_{\mu;\nu}$ and $*F_{\mu\nu} = M_{\nu;\mu} - M_{\mu;\nu}$, so that these may be directly related to one another by $M_{\nu;\mu} - M_{\mu;\nu} = \tfrac{1}{2!}\varepsilon_{\delta\gamma\mu\nu}(A^{\gamma;\delta} - A^{\delta;\gamma})$. In effect, in Minkowski spacetime with $g_{\mu\nu} = \eta_{\mu\nu}$, this cuts in half, the degree of freedom which $J^\nu$, $P^\nu$, and $A_\nu, M_\nu$ would otherwise have if $F_{\mu\nu}$ and $*F_{\mu\nu}$ were to be totally independent fields tensors.[*] It stands to reason, therefore, when it comes to $\psi_e$ and $\psi_m$, that we should eventually come across some freedom-reducing constraint which makes also these – not totally-independent of one another – but related

---

[*] It is also worth noting that in a gravitational field, $g_{\mu\nu} \neq \eta_{\mu\nu}$, due to the covariant $*F^{\sigma\tau} \equiv \tfrac{1}{2!}\varepsilon^{\delta\gamma\sigma\tau}F_{\delta\gamma}$, the $J^\nu$, $P^\nu$, as well as $A_\nu, M_\nu$ acquire additional independence from one another due to the ten independent components of the symmetric metric tensor. Thanks to Ken S. Tucker who first pointed this out on sci.physics.



in some manner. (4.11) and (4.12) now tell us what this constraint might be. In (4.11), we learn that $\psi_e = \psi_R$ and in (4.12) that $\psi_m = \psi_L$ (or vice versa). That is, duality constrains unmixed electric and magnetic monopole currents so that when all is said and done, *the fermions which constitute these currents end up being the opposite chirality states of the same observed fermions, rather than distinctly independent charges*. Note, even if we imposed some other constraint than $|c_A| = |c_V|$, that equation (4.5), $c_A(Q_e) = -c_A(Q_m)$ ensures that $\psi_e$ and $\psi_m$ will nevertheless be different chiral manifestations of the same Fermion charge, unless $J^\nu$, $P^\nu$ are each chiral symmetric,. Thus, we find, possibly, a very deep connection between electric and magnetic monopoles, and chiral symmetries and asymmetries.

Further, with the constraint $|c_A| = |c_V|$, i.e., if $\psi_e = \psi_R$ and $\psi_m = \psi_L$, then each of $\psi_e$ and $\psi_m$ is a two component Dirac spinor. Such spinors, of course are *massless*. Yet, as soon as we break symmetry by setting $J_{em}{}^\mu = \bar\psi \gamma^\mu Q \psi \equiv J^{\mu\prime} = J^\mu + P^\mu = \bar\psi_R \gamma^\mu Q \psi_R + \bar\psi_L \gamma^\mu Q \psi_L$ in (4.10), we have created a four-component spinor $\psi$ out of $\psi_e = \psi_R$ and $\psi_m = \psi_L$. *This spinor now has mass because it can be overtaken by a Lorentz transformation!* In the Higgs Goldstone mechanism, a massless vector boson with two transverse polarization states acquires a mass when it swallows up a Goldstone scalar to gain a third longitudinal polarization. It may well be that here, the $\psi_e = \psi_R$ and $\psi_m = \psi_L$ are *swallowing up one another* to go from being two two-component fermions which are each massless, to being a single four-component Fermion which is massive. In short, the particular constraint $|c_A| = |c_V|$ which leads to $\psi_e = \psi_R$ and $\psi_m = \psi_L$ may plant the seeds of a mechanism by which Fermion mass can be generated.

This may also give us a better understanding of why the *weak* interaction is V-A rather than chiral symmetric. We know that weak parity violation is the hand that nature has dealt us for weak interactions *at low energy*, but in the almost half a century since Lee and Yang first discovered this, we still don't know *why* nature serves up this seeming complication. If duality symmetry leads generally to $\psi_e = \psi_R$ and $\psi_m = \psi_L$ for Fermions before their currents are mixed to become observable according to an equation like (4.10), then this means that for any interaction, nature may well begin with two separate Lagrangians, one for right-handed and "chromo-electrically" charged fermions $\psi_e = \psi_R$, the other left-handed and "chromo-magnetically" charged fermions $\psi_m = \psi_L$ (again, or vice versa). In electromagnetic and strong interactions, we observe a Lagrangian *following* mixing (2.2) and symmetry breaking (2.5). But in SU(2)$_L$ weak interactions, we observe a Lagrangian *before* mixing (2.2) and symmetry breaking (2.5) with a "weak electric" charge $g_{we}$ coupled to a V-A current observable at energies in the 100 GeV range (M$_{w,z}$). This could mean, however, that there is also an SU(2)$_R$ interaction which many suspect exists (see, e.g., [2] at section 12.2) involving a much larger "weak magnetic" charge $g_{wm}$ coupled to a V+A current observable only at much higher energies. In this event, the Dirac Quantization Condition for unit *weak electric and magnetic charges* now becomes $g_{we} \cdot g_{wm} = n \cdot 2\pi\hbar c$.

In short, the possibility that electric and magnetic monopoles may generally be related to chiral fermion states according to $\psi_e = \psi_R$ and $\psi_m = \psi_L$ (or vice versa) may explain several of nature's deepest mysteries about the nature of chirality and, perhaps, at the same time, lead to a mechanism for generating fermion masses.



## 5. Calculation of Vertex Factors for the Magnetic Monopole Interaction

Based on the development in section 4, we now lay the foundation for calculating magnetic monopole widths and cross sections, by calculating the magnetic monopole vertex factors for the leptons and quarks.

In [1], at (6.7), the author deduced that for $a_e' = 1/137.036$ at low energy,

$$\sin^2 \alpha = 2.131 \times 10^{-4}; \quad \cos^2 \alpha = .9998. \tag{5.1}$$

Even with the electromagnetic running coupling rising to $a_e' \sim 1/126$ near 2TeV, see discussion in [1] following (5.23), it is clear that the magnetic monopole current (4.13) will be very-heavily biased toward left-handed chirality (or right handed chirality if nature makes the opposite choice), by a factor of about 4000 to 1. So, when and if the magnetic monopole interaction $-g_m' P^{\mu'} M_\mu'$ is actually observed, it should be clear how to properly assign the V-A and the V+A currents.

At this point, let us rewrite (4.13) as:

$$P^{\mu'} = \overline{\psi}_L \gamma^\mu Q \psi_L - \sin^2 \alpha \cdot \overline{\psi} \gamma^\mu Q \psi = \overline{\psi} \gamma^\mu \tfrac{1}{2}(1 - \gamma^5) Q \psi - \sin^2 \alpha \cdot \overline{\psi} \gamma^\mu Q \psi$$
$$= \overline{\psi} \gamma^\mu [(\tfrac{1}{2} - \sin^2 \alpha) Q - \tfrac{1}{2} \gamma^5 Q] \psi \equiv \overline{\psi} \gamma^\mu (\tfrac{1}{2} c_V(P) - \tfrac{1}{2} c_A(P) \gamma^5) \psi \tag{5.2}$$

from which we can identify the chiral vertex factors for the magnetic monopole current (the final approximations in (5.3), (5.5) and (5.6) are due to (5.1)):

$$c_V(P) = Q(1 - 2\sin^2 \alpha) = Q(\cos^2 \alpha - \sin^2 \alpha) = Q \cdot \cos 2\alpha \cong Q. \tag{5.3}$$

$$c_A(P) = Q. \tag{5.4}$$

$$c_R(P) \equiv c_V(P) - c_A(P) = Q(\cos 2\alpha - 1) = Q(\cos^2 \alpha - \sin^2 \alpha - 1) = -2Q \cdot \sin^2 \alpha \cong 0. \tag{5.5}$$

$$c_L(P) \equiv c_V(P) + c_A(P) = Q(\cos 2\alpha + 1) = Q(\cos^2 \alpha - \sin^2 \alpha + 1) = 2Q \cdot \cos^2 \alpha \cong 2Q. \tag{5.6}$$

From (3.4), we take $\overline{\psi} \gamma^\mu (\tfrac{1}{2}(1-\gamma^5) I_3 - \sin^2 \theta_W \cdot Q) \psi \equiv \overline{\psi} \gamma^\mu \tfrac{1}{2}(c_V(Z) - c_A(Z)\gamma^5) \psi$, and can thereby identify the usual electroweak Z-vertexes:

$$c_V(Z) = I_3 - 2\sin^2 \theta_W \cdot Q. \tag{5.7}$$

$$c_A(Z) = I_3. \tag{5.8}$$

$$c_R(Z) \equiv c_V(Z) - c_A(Z) = -2Q \cdot \sin^2 \theta_W. \tag{5.8}$$

$$c_L(Z) \equiv c_V(Z) + c_A(Z) = 2(I_3 - Q \cdot \sin^2 \theta_W). \tag{5.9}$$



Now, to derive strictly numeric values for all the vertex factors for all Fermions, from equation (6.6) of [1], we can deduce the complexion $\alpha$ based on an estimated running coupling $a_e' \sim 1/126$ at around 2.5 TeV, according to (using the negative root):

$$\sin^2 \alpha = \frac{1 \mp \sqrt{1-16 a_e'^2}}{2} = 2.52 \times 10^{-4}; \quad \cos^2 \alpha = .9997. \tag{5.10}$$

Again, $\sin^2 \alpha$ is still so small that the approximations in (5.3), (5.5) and (5.6) remain useful and simplifying. Additionally, we may employ $\sin^2 \theta_W (M_Z) = .23120$ from [3], because the detailed calculations to follow will be at $\sqrt{s} = M_Z$. Then, using the known values of Q and $I_3$, we may write, for each fermion:

$$\nu, \nu_\mu, \nu_\tau = \left| \begin{array}{l} Q=0, I_3 = \tfrac{1}{2}, \\ c_V(P) = 0, c_A(P) = 0, c_R(P) = 0, c_L(P) = 0 \\ c_V(Z) = \tfrac{1}{2}, c_A(Z) = \tfrac{1}{2}, c_R(Z) = 0, c_L(Z) = 1 \end{array} \right\rangle. \tag{5.11}$$

$$e, \mu, \tau = \left| \begin{array}{l} Q=-1, I_3 = -\tfrac{1}{2}, \\ c_V(P) = -\cos 2\alpha \cong -1, c_A(P) = -1, c_R(P) = 2\sin^2 \alpha \cong 0, c_L(P) = -2\cos^2 \alpha \cong -2 \\ c_V(Z) = -\tfrac{1}{2} + 2\sin^2 \theta_W = -.0376, c_A(Z) = -\tfrac{1}{2}, c_R(Z) = 2\sin^2 \theta_W, c_L(Z) = -\cos 2\theta_W \end{array} \right\rangle. \tag{5.12}$$

$$u, c, t = \left| \begin{array}{l} Q=\tfrac{2}{3}, I_3 = \tfrac{1}{2} \\ c_V(P) = \tfrac{2}{3}\cos 2\alpha \cong \tfrac{2}{3}, c_A(P) = \tfrac{2}{3}, c_R(P) = -\tfrac{4}{3}\sin^2 \alpha \cong 0, c_L(P) = \tfrac{4}{3}\cos^2 \alpha \cong \tfrac{4}{3} \\ c_V(Z) = \tfrac{1}{2} - \tfrac{4}{3}\sin^2 \theta_W = .1917, c_A(Z) = \tfrac{1}{2}, c_R(Z) = -\tfrac{4}{3}\sin^2 \theta_W, c_L(Z) = 1 - \tfrac{4}{3}\sin^2 \theta_W \end{array} \right\rangle. \tag{5.13}$$

$$d, s, b = \left| \begin{array}{l} Q=-\tfrac{1}{3}, I_3 = -\tfrac{1}{2} \\ c_V(P) = -\tfrac{1}{3}\cos 2\alpha \cong -\tfrac{1}{3}, c_A(P) = -\tfrac{1}{3}, c_R(P) = \tfrac{2}{3}\sin^2 \alpha \cong 0, c_L(P) = -\tfrac{2}{3}\cos^2 \alpha \cong -\tfrac{2}{3} \\ c_V(Z) = -\tfrac{1}{2} + \tfrac{2}{3}\sin^2 \theta_W = -.3459, c_A(Z) = -\tfrac{1}{2}, c_R(Z) = \tfrac{2}{3}\sin^2 \theta_W, c_L(Z) = -1 + \tfrac{2}{3}\sin^2 \theta_W \end{array} \right\rangle. \tag{5.14}$$

All of the above will be crucial ingredients to helping us deduce the widths and cross sections of the magnetic monopole current interaction.

## 6. Width Calculations for the Electroweak Neutral Current and Magnetic Monopole Vector Bosons

The first step on the path to calculating cross sections is to calculate the partial and full widths for the $M_\mu'$ vector boson which, as shown just following (8.14) in [1], has a mass of about 2.35 TeV when observed at low-TeV energy. We recall also from [1] that Mass ($M^{\mu'}$) =



$\frac{1}{2} v g_m{}'$ in general, and that the 2.35 TeV mass comes about from *assuming* that $v = v_F = 246.220$ GeV, see [1] just following (8.13).

In general, the partial width for the decay of a vector boson X into two spin ½ fermions $f_1$ and $\bar{f}_2$, where $M_X \gg m_f$ and so the fermion masses can be neglected, is (see, e.g., [4], equation (13.43)):

$$\Gamma(X \to f_1 \bar{f}_2) = \frac{g_X^2}{48\pi}(c_V^2 + c_A^2)M_X \tag{6.1}$$

In (5.11) to (5.14), we already have calculated the necessary vertex factors for all of the known fermions. As a point of departure, let us first see how the widths are calculated for the electroweak $Z^\mu$ vector boson. The, we shall follow an identical set of steps to calculate the widths for the $M_\mu{}'$ vector boson which mediates the observable magnetic monopole interactions.

For the $Z^\mu$, as a "warm up" exercise, we shall use $M_Z = 91.1876\,GeV$ [5], $g_z = \frac{g_w}{\cos\theta_W} = \frac{g_e{}'}{\sin\theta_W \cos\theta_W}$, $a_e{}' = \frac{g_e{}'^2}{4\pi\hbar c} = \frac{1}{137.036}$, and as set forth following (5.10), $\sin^2\theta_W(M_Z) = .23120$.

For the neutrinos, we use (5.11) to deduce:

$$\Gamma(Z \to \bar{\nu}\nu) = \frac{g_z^2}{48\pi}(c_V^\nu(Z)^2 + c_A^\nu(Z)^2)M_Z = \frac{g_w^2}{48\pi\cos^2\theta_W}\frac{1}{2}M_Z = .156\,GeV \tag{6.2}$$

For the electrons, using (5.12), we find:

$$\Gamma(Z \to \bar{e}e) = \frac{g_z^2}{48\pi}(c_V^e(Z)^2 + c_A^e(Z)^2)M_Z = \frac{g_w^2}{48\pi\cos^2\theta_W}(.2514)M_Z = .078\,GeV. \tag{6.3}$$

For the up quarks, using (5.13), and multiplying by 3 for three colors, we find:

$$\Gamma(Z \to \bar{u}u) = \frac{g_z^2}{48\pi}(c_V^u(Z)^2 + c_A^u(Z)^2)M_Z = \frac{g_w^2}{48\pi\cos^2\theta_W}(.2869)M_Z = .095\,GeV \Rightarrow \times 3 = .267\,GeV. \tag{6.4}$$

For the down quarks, using (5.14), and again multiplying by 3 for color, we find:

$$\Gamma(Z \to \bar{d}d) = \frac{g_z^2}{48\pi}(c_V^d(Z)^2 + c_A^d(Z)^2)M_Z = \frac{g_w^2}{48\pi\cos^2\theta_W}(.3697)M_Z = .123\,GeV \Rightarrow \times 3 = .345\,GeV. \tag{6.5}$$

Summing the widths in (6.2) to (6.5), and multiplying the entire result by 3 for three generations, we arrive at the full width:

$$\Gamma_Z = 2.538\,GeV \tag{6.6}$$



which contrasts nicely to the experimental value $\Gamma(Z) = 2.4952\, GeV$.

Now, we follow the same path for the $M_\mu'$. Here, we shall estimate $M_M \equiv M(M_\mu') \cong 2.35\, TeV$, $a_e' \cong \frac{1}{126}$ near 2 TeV, $\sin^2\alpha = \frac{1 \mp \sqrt{1-16a_e'^2}}{2} = 2.52 \times 10^{-4}$ (see (5.10)), and, since $2a_m' = \frac{1}{2a_e'} = \frac{g_m'^2}{2\pi\hbar c} = \frac{2\pi\hbar c}{g_e'^2}$, see (5.22) in [1], $g_m'^2 = \frac{\pi\hbar c}{a_e'} = 395.841$.

For the neutrinos, we use (5.11) to deduce:

$$\Gamma(M \to \bar{\nu}\nu) = \frac{g_m'^2}{48\pi}(c_V(P)^2 + c_A(P)^2)M_M = \frac{395.841}{48\pi}(0)M_M = 0\, GeV. \tag{6.7}$$

This is zero because the neutrino carries no electric or magnetic charge. For the electrons, using (5.12), we find:

$$\Gamma(M \to \bar{e}e) = \frac{g_m'^2}{48\pi}(c_V^e(P)^2 + c_A^e(P)^2)M_M = \frac{395.841}{48\pi}1.999 \cdot (2.35\, TeV) = 12.33\, TeV. \tag{6.8}$$

For the up quarks, using (5.13), and multiplying by 3 for three colors, we find:

$$\Gamma(M \to \bar{u}u) = \frac{g_m'^2}{48\pi}(c_V^u(P)^2 + c_A^u(P)^2)M_M = \frac{395.841}{48\pi}.888(2.35\, TeV) = 5.48\, TeV \Rightarrow \times 3 = 16.44\, TeV. \tag{6.9}$$

For the down quarks, using (5.14), and again multiplying by 3 for color, we find:

$$\Gamma(M \to \bar{d}d) = \frac{g_m'^2}{48\pi}(c_V^d(P)^2 + c_A^d(P)^2)M_M = \frac{395.841}{48\pi}.222(2.35\, TeV) = 1.37\, TeV \Rightarrow \times 3 = 4.11\, TeV. \tag{6.10}$$

Summing the widths in (6.7) to (6.10), and multiplying the entire result by 3 for three generations, we arrive at an estimated full width:

$$\Gamma_M = 98.64\, TeV, \tag{6.11}$$

which, of course, corresponds to an exceptionally short lifetime. Note, this was based on $M_M \equiv M(M_\mu') \cong 2.35\, TeV$ which in turn assumes that the vev for duality $v = v_F = 246.220$ GeV. For different $v$, the width in (6.11) would vary linearly with $v$, that is, $\Gamma_M = 98.64\, TeV \times \frac{v}{v_F}$. Similarly, $M_M = 2.35\, TeV \times \frac{v}{v_F}$.

Note that the estimated full width for the $M_\mu'$ is larger than it 2.35 TeV mass by a factor of 41.97. In contrast, the approximate full width for the $Z^\mu$ is smaller than its mass by a factor of about 36.54. This is because for the $Z^\mu$, the small charge $g_z^2 = \frac{.3965}{.7688} = .5157$ was the



dominant factor in the width formulae, while for the $M_\mu'$, the huge charge $g_m'^2 = \frac{\pi \hbar c}{a_e'} = 395.841$ predominates. In choosing $a_e' \cong \frac{1}{126}$ near 2 TeV, we already account for perturbative charge screening in the small electromagnetic coupling $a_e'$. From there, the Dirac quantization condition for unit charge immediately sets $g_m'^2 = \frac{\pi \hbar c}{a_e'} = 395.841$, which *already incorporates a perturbative result insofar as perturbation theory has been applied to find* $a_e' \cong \frac{1}{126}$ *near 2 TeV*. This is a good example of how the Dirac quantization condition (2.9) serves to provide a very large coupling >>1 which is nevertheless perturbatively valid, because of the fact that this coupling is an "inverse" of a very small coupling <<1.

## 7. Development of Cross Section Formulas for Magnetic Monopole Interactions

At this point, let turn to the Lagrangian (8.18) from [1], which we reproduce below, including the full width calculated in (6.11), as such:

$$L_{gJB}' = -g_e' J^{\mu'} \frac{-g_{\mu\lambda}}{p^\tau p_\tau} J^{\lambda'} - g_m' P^{\mu'} \frac{-g_{\mu\lambda} + p_\mu p_\lambda / M_M^2}{p^\tau p_\tau - M_M^2 + iM_M \Gamma_M} P^{\lambda'}. \tag{7.1}$$

Let us also contrast the similar Lagrangian for the neutral sector of electroweak theory, given by:

$$L_{gJB}' = -g_e' J^{\mu'} \frac{-g_{\mu\lambda}}{p^\tau p_\tau} J^{\lambda'} - g_z J_Z^\mu \frac{-g_{\mu\lambda} + p_\mu p_\lambda / M_Z^2}{p^\tau p_\tau - M_Z^2 + iM_Z \Gamma_Z} J_Z^\lambda. \tag{7.2}$$

From the above, it is straightforward to identify the three invariant amplitudes:

$$\mathfrak{M}_A = -g_e'^2 J^{\mu'} \frac{g_{\mu\lambda}}{p^\tau p_\tau} J^{\lambda'}. \tag{7.3}$$

$$\mathfrak{M}_Z = -g_z^2 J_Z^\mu \frac{-g_{\mu\lambda} + p_\mu p_\lambda / M_Z^2}{p^\tau p_\tau - M_Z^2 + iM_Z \Gamma_Z} J_Z^\lambda \tag{7.4}$$

$$\mathfrak{M}_M = -g_m'^2 P^{\mu'} \frac{-g_{\mu\lambda} + p_\mu p_\lambda / M_M^2}{p^\tau p_\tau - M_M^2 + iM_M \Gamma_M} P^{\lambda'}. \tag{7.5}$$

We will want to calculate scattering cross sections for both $e^+ e^- \to \mu^+ \mu^-$ and $e^+ e^- \to q\bar{q}$ via all three of the photon $A^\mu$, the weak neutral current $Z^\mu$, and the $M^\mu$ which mediates magnetic monopole interactions.



First, we rewrite (7.3) to (7.5) such that the "$e^+e^- \to$" is represented by an $\bar{e}\gamma^\mu e$ vertex on the right side of the propagator and the "$\to \bar{f}f$" is represented in general terms by an $\bar{f}\gamma^\mu f$ vertex to the left of the propagator. The electron of course has electric (and magnetic) charge Q = -1, so we set $J^{\lambda\prime} = -\bar{e}\gamma^\lambda e$, $J^{\mu\prime} = \bar{f}\gamma^\mu Q_f f$. Additionally, to conserve the momentum during scattering from initial (A,B) to final (C,D) states, we set $p_A^\mu + p_B^\mu = p_C^\mu + p_D^\mu$, and thus in (7.3) to (7.5), may set $J_Z^\lambda = \tfrac{1}{2}\bar{e}\gamma^\lambda(c_V^e(Z) - c_A^e(Z)\gamma^5)e$, $J_Z^\mu = \tfrac{1}{2}\bar{f}\gamma^\mu(c_V^f(Z) - c_A^f(Z)\gamma^5)f$, $P^{\lambda\prime} = \tfrac{1}{2}\bar{e}\gamma^\lambda(c_V^e(P) - c_A^e(P)\gamma^5)e$, and $P^{\mu\prime} = \tfrac{1}{2}\bar{f}\gamma^\mu(c_V^f(P) - c_A^f(P)\gamma^5)f$, to obtain:

$$\mathfrak{M}_A = +g_e^{\prime 2}(\bar{f}\gamma^\mu Q_f f)\frac{g_{\mu\lambda}}{p^\tau p_\tau}(\bar{e}\gamma^\lambda e). \tag{7.6}$$

$$\mathfrak{M}_Z = -\tfrac{1}{4}g_z^2 [\bar{f}\gamma^\mu(c_V^f(Z) - c_A^f(Z)\gamma^5)f]\frac{-g_{\mu\lambda} + p_\mu p_\lambda/M_Z^2}{p^\tau p_\tau - M_Z^2 + iM_Z\Gamma_Z}[\bar{e}\gamma^\lambda(c_V^e(Z) - c_A^e(Z)\gamma^5)e] \tag{7.7}$$

$$\mathfrak{M}_M = -\tfrac{1}{4}g_m^{\prime 2} [\bar{f}\gamma^\mu(c_V^f(P) - c_A^f(P)\gamma^5)f]\frac{-g_{\mu\lambda} + p_\mu p_\lambda/M_M^2}{p^\tau p_\tau - M_M^2 + iM_M\Gamma_M}[\bar{e}\gamma^\lambda(c_V^e(P) - c_A^e(P)\gamma^5)e]. \tag{7.8}$$

Now we use the identity $(c_V - c_A\gamma^5) = (c_V - c_A)\tfrac{1}{2}(1+\gamma^5) + (c_V + c_A)\tfrac{1}{2}(1-\gamma^5)$ and the definitions $c_R \equiv c_V - c_A$ and $c_L \equiv c_V + c_A$ to rewrite the above in terms of left- and right-handed spinors as:

$$\mathfrak{M}_A = g_e^{\prime 2}\frac{g_{\mu\lambda}}{p^\tau p_\tau}Q_f[(\bar{f}_R\gamma^\mu f_R)(\bar{e}_R\gamma^\lambda e_R) + (\bar{f}_L\gamma^\mu f_L)(\bar{e}_L\gamma^\lambda e_L) + (\bar{f}_L\gamma^\mu f_L)(\bar{e}_R\gamma^\lambda e_R) + (\bar{f}_R\gamma^\mu f_R)(\bar{e}_L\gamma^\lambda e_L)]. \tag{7.9}$$

$$\mathfrak{M}_Z = -\tfrac{1}{4}g_z^2\frac{-g_{\mu\lambda} + p_\mu p_\lambda/M_Z^2}{p^\tau p_\tau - M_Z^2 + iM_Z\Gamma_Z}\times\begin{bmatrix}c_R^f(Z)c_R^e(Z)(\bar{f}_R\gamma^\mu f_R)(\bar{e}_R\gamma^\lambda e_R) + c_L^f(Z)c_L^e(Z)(\bar{f}_L\gamma^\mu f_L)(\bar{e}_L\gamma^\lambda e_L) \\ + c_L^f(Z)c_R^e(Z)(\bar{f}_L\gamma^\mu f_L)(\bar{e}_R\gamma^\lambda e_R) + c_R^f(Z)c_L^e(Z)(\bar{f}_R\gamma^\mu f_R)(\bar{e}_L\gamma^\lambda e_L)\end{bmatrix} \tag{7.10}$$

$$\mathfrak{M}_M = -\tfrac{1}{4}g_m^{\prime 2}\frac{-g_{\mu\lambda} + p_\mu p_\lambda/M_M^2}{p^\tau p_\tau - M_M^2 + iM_M\Gamma_M}\times\begin{bmatrix}c_R^f(P)c_R^e(P)(\bar{f}_R\gamma^\mu f_R)(\bar{e}_R\gamma^\lambda e_R) + c_L^f(P)c_L^e(P)(\bar{f}_L\gamma^\mu f_L)(\bar{e}_L\gamma^\lambda e_L) \\ + c_L^f(P)c_R^e(P)(\bar{f}_L\gamma^\mu f_L)(\bar{e}_R\gamma^\lambda e_R) + c_R^f(P)c_L^e(P)(\bar{f}_R\gamma^\mu f_R)(\bar{e}_L\gamma^\lambda e_L)\end{bmatrix}. \tag{7.11}$$

Now, we turn to calculate the full amplitude $\mathfrak{M}_A + \mathfrak{M}_Z + \mathfrak{M}_M$ for each of the four helicity configurations in (7.9) through (7.11). The development to follow parallels [4], section 13.6. For the helicity states $[(e_R f_R \to e_R f_R) \Leftrightarrow (e_R\bar{e}_L \to f_R\bar{f}_L)]$, $[(e_L f_L \to e_L f_L) \Leftrightarrow (e_L\bar{e}_R \to f_L\bar{f}_R)]$, $[(e_R f_L \to e_R f_L) \Leftrightarrow (e_R\bar{e}_L \to f_L\bar{f}_R)]$, and $[(e_L f_R \to e_L f_R) \Leftrightarrow (e_L\bar{e}_R \to f_R\bar{f}_L)]$, we obtain:

$$\mathfrak{M}_A + \mathfrak{M}_Z + \mathfrak{M}_M (e_R\bar{e}_L \to f_R\bar{f}_L) = \frac{g_e^{\prime 2}}{p^\tau p_\tau}[Q_f g_{\mu\lambda} + r_Z c_R^f(Z)c_R^e(Z) + r_M c_R^f(P)c_R^e(P)](\bar{f}_R\gamma^\mu f_R)(\bar{e}_R\gamma^\lambda e_R) \tag{7.12}$$



$$\mathfrak{M}_A + \mathfrak{M}_Z + \mathfrak{M}_M \ (e_L \bar{e}_R \to f_L \bar{f}_R) = \frac{g_e^{'2}}{p^\tau p_\tau} \left[ Q_f g_{\mu\lambda} + r_Z c_L^f(Z) c_L^e(Z) + r_M c_L^f(P) c_L^e(P) \right] (\bar{f}_L \gamma^\mu f_L)(\bar{e}_L \gamma^\lambda e_L), \quad (7.13)$$

$$\mathfrak{M}_A + \mathfrak{M}_Z + \mathfrak{M}_M \ (e_R \bar{e}_L \to f_L \bar{f}_R) = \frac{g_e^{'2}}{p^\tau p_\tau} \left[ Q_f g_{\mu\lambda} + r_Z c_L^f(Z) c_R^e(Z) + r_M c_L^f(P) c_R^e(P) \right] (\bar{f}_L \gamma^\mu f_L)(\bar{e}_R \gamma^\lambda e_R), \quad (7.14)$$

$$\mathfrak{M}_A + \mathfrak{M}_Z + \mathfrak{M}_M \ (e_L \bar{e}_R \to f_R \bar{f}_L) = \frac{g_e^{'2}}{p^\tau p_\tau} \left[ Q_f g_{\mu\lambda} + r_Z c_R^f(Z) c_L^e(Z) + r_M c_R^f(P) c_L^e(P) \right] (\bar{f}_R \gamma^\mu f_R)(\bar{e}_L \gamma^\lambda e_L), \quad (7.15)$$

where for all of the above, we have defined the ratios:

$$r_Z \equiv -\tfrac{1}{4} g_z^2 \frac{-g_{\mu\lambda} + p_\mu p_\lambda / M_Z^2}{p^\tau p_\tau - M_Z^2 + iM_Z \Gamma_Z} \frac{p^\tau p_\tau}{g_e^{'2}}. \quad (7.16)$$

and:

$$r_M \equiv -\tfrac{1}{4} g_m^{'2} \frac{-g_{\mu\lambda} + p_\mu p_\lambda / M_M^2}{p^\tau p_\tau - M_M^2 + iM_M \Gamma_M} \frac{p^\tau p_\tau}{g_e^{'2}}. \quad (7.17)$$

We may ignore the incident lepton masses, so (7.12) through (7.15) simplify to:

$$\mathfrak{M}_A + \mathfrak{M}_Z + \mathfrak{M}_M \ (e_R \bar{e}_L \to f_R \bar{f}_L) = \frac{g_e^{'2}}{p^\tau p_\tau} \left[ Q_f + r_Z c_R^f(Z) c_R^e(Z) + r_M c_R^f(P) c_R^e(P) \right] (\bar{f}_R \gamma^\mu f_R)(\bar{e}_R \gamma_\mu e_R), \quad (7.18)$$

$$\mathfrak{M}_A + \mathfrak{M}_Z + \mathfrak{M}_M \ (e_L \bar{e}_R \to f_L \bar{f}_R) = \frac{g_e^{'2}}{p^\tau p_\tau} \left[ Q_f + r_Z c_L^f(Z) c_L^e(Z) + r_M c_L^f(P) c_L^e(P) \right] (\bar{f}_L \gamma^\mu f_L)(\bar{e}_L \gamma_\mu e_L), \quad (7.19)$$

$$\mathfrak{M}_A + \mathfrak{M}_Z + \mathfrak{M}_M \ (e_R \bar{e}_L \to f_L \bar{f}_R) = \frac{g_e^{'2}}{p^\tau p_\tau} \left[ Q_f + r_Z c_L^f(Z) c_R^e(Z) + r_M c_L^f(P) c_R^e(P) \right] (\bar{f}_L \gamma^\mu f_L)(\bar{e}_R \gamma_\mu e_R), \quad (7.20)$$

$$\mathfrak{M}_A + \mathfrak{M}_Z + \mathfrak{M}_M \ (e_L \bar{e}_R \to f_R \bar{f}_L) = \frac{g_e^{'2}}{p^\tau p_\tau} \left[ Q_f + r_Z c_R^f(Z) c_L^e(Z) + r_M c_R^f(P) c_L^e(P) \right] (\bar{f}_R \gamma^\mu f_R)(\bar{e}_L \gamma_\mu e_L), \quad (7.21)$$

with

$$r_Z \equiv \tfrac{1}{4} \frac{g_z^2}{p^\tau p_\tau - M_Z^2 + iM_Z \Gamma_Z} \frac{p^\tau p_\tau}{g_e^{'2}}. \quad (7.22)$$

and:

$$r_M \equiv \tfrac{1}{4} \frac{g_m^{'2}}{p^\tau p_\tau - M_M^2 + iM_M \Gamma_M} \frac{p^\tau p_\tau}{g_e^{'2}}. \quad (7.23)$$



Next, in preparation for cross section calculations, we need to deduce $|\mathcal{M}_A + \mathcal{M}_Z + \mathcal{M}_M|^2$ For $(e_R \bar{e}_L \to f_R \bar{f}_L)$, $(e_L \bar{e}_R \to f_L \bar{f}_R)$, $(e_R \bar{e}_L \to f_L \bar{f}_R)$, and $(e_L \bar{e}_R \to f_R \bar{f}_L)$ decays respectively, and for high energy with $s \equiv p^\tau p_\tau$, also using $g_e'^2 = 4\pi a_e'$, we obtain:

$$|\mathcal{M}_A + \mathcal{M}_Z + \mathcal{M}_M|^2 \cong g_e'^4 \left| Q_f + r_Z c_R^f(Z) c_R^e(Z) + r_M c_R^f(P) c_R^e(P) \right|^2 \frac{u^2}{s^2}$$
$$\cong 4\pi^2 a_e'^2 \left| Q_f + r_Z c_R^f(Z) c_R^e(Z) + r_M c_R^f(P) c_R^e(P) \right|^2 \left[1 + 2\cos\theta + \cos^2\theta\right], \quad (7.24)$$

$$|\mathcal{M}_A + \mathcal{M}_Z + \mathcal{M}_M|^2 \cong g_e'^4 \left| Q_f + r_Z c_L^f(Z) c_L^e(Z) + r_M c_L^f(P) c_L^e(P) \right|^2 \frac{u^2}{s^2}$$
$$\cong 4\pi^2 a_e'^2 \left| Q_f + r_Z c_L^f(Z) c_L^e(Z) + r_M c_L^f(P) c_L^e(P) \right|^2 \left[1 + 2\cos\theta + \cos^2\theta\right], \quad (7.25)$$

$$|\mathcal{M}_A + \mathcal{M}_Z + \mathcal{M}_M|^2 \cong g_e'^4 \left| Q_f + r_Z c_L^f(Z) c_R^e(Z) + r_M c_L^f(P) c_R^e(P) \right|^2 \frac{t^2}{s^2}$$
$$\cong 4\pi^2 a_e'^2 \left| Q_f + r_Z c_L^f(Z) c_R^e(Z) + r_M c_L^f(P) c_R^e(P) \right|^2 \left[1 - 2\cos\theta + \cos^2\theta\right], \quad (7.26)$$

$$|\mathcal{M}_A + \mathcal{M}_Z + \mathcal{M}_M|^2 \cong g_e'^4 \left| Q_f + r_Z c_R^f(Z) c_L^e(Z) + r_M c_R^f(P) c_L^e(P) \right|^2 \frac{t^2}{s^2}$$
$$\cong 4\pi^2 a_e'^2 \left| Q_f + r_Z c_R^f(Z) c_L^e(Z) + r_M c_R^f(P) c_L^e(P) \right|^2 \left[1 - 2\cos\theta + \cos^2\theta\right], \quad (7.27)$$

where we have used the angular momentum $L_f^{\mu\nu} L_{e\mu\nu} = \frac{1}{4} \sum_{spins} (\bar{f}\gamma^\mu f)(\bar{f}\gamma^\nu f)(\bar{e}\gamma_\mu e)(\bar{e}\gamma_\nu e)$ and the trace theorems of the Dirac matrices to obtain, for high energy:

$$(\bar{f}_R \gamma^\mu f_R)(\bar{f}_R \gamma^\nu f_R)(\bar{e}_R \gamma_\mu e_R)(\bar{e}_R \gamma_\nu e_R) = (\bar{f}_L \gamma^\mu f_L)(\bar{f}_L \gamma^\nu f_L)(\bar{e}_L \gamma_\mu e_L)(\bar{e}_L \gamma_\nu e_L) \cong u^2, \quad (7.28)$$

$$(\bar{f}_L \gamma^\mu f_L)(\bar{f}_L \gamma^\nu f_L)(\bar{e}_R \gamma_\mu e_R)(\bar{e}_R \gamma_\nu e_R) = (\bar{f}_R \gamma^\mu f_R)(\bar{f}_R \gamma^\nu f_R)(\bar{e}_L \gamma_\mu e_L)(\bar{e}_L \gamma_\nu e_L) \cong t^2. \quad (7.29)$$

We have also used the Mandelstam variables:

$$s = 4(p^2 + m^2), \quad t = -2p^2(1 - \cos\theta), \quad u = -2p^2(1 + \cos\theta), \quad (7.30)$$

hence, also for high energy:

$$\frac{t^2}{s^2} = \frac{4p^4(1 - 2\cos\theta + \cos^2\theta)}{16(p^4 + p^2 m^2 + m^4)} \cong \frac{(1 - 2\cos\theta + \cos^2\theta)}{4}, \quad (7.31)$$

$$\frac{u^2}{s^2} = \frac{4p^4(1 + 2\cos\theta + \cos^2\theta)}{16(p^4 + p^2 m^2 + m^4)} \cong \frac{(1 + 2\cos\theta + \cos^2\theta)}{4}. \quad (7.32)$$



Now, we sum the spin states. For the vertex factors, it is helpful to use $c_R \equiv c_V - c_A$ and $c_L \equiv c_V + c_A$, so that $c_V = \frac{1}{2}(c_R + c_L)$; $c_A = \frac{1}{2}(c_L - c_R)$, as well as $c_R^2 = c_V^2 + c_A^2 - 2c_V c_A$; $c_L^2 = c_V^2 + c_A^2 + 2c_V c_A$; $c_R^2 + c_L^2 = 2(c_V^2 + c_A^2)$; and $c_R^2 - c_L^2 = -4 c_V c_A$.

From (7.24) through (7.27), we may then deduce:

$$\sum_{spins} |\mathfrak{M}_A + \mathfrak{M}_A + \mathfrak{M}_M|^2 = 16\pi^2 a_e'^2 \left[ A_0 (1 + \cos^2 \theta) + A_1 \cos \theta \right], \tag{7.33}$$

where:

$$A_0 = Q_f^2 + A_0(Z) + A_0(P) + A_0(Z \Leftrightarrow P) \tag{7.34}$$

$$A_0(Z) = \tfrac{1}{2} Q_f \operatorname{Re}(r_Z) \left( c_R^f(Z) + c_L^f(Z) \right)\left( c_R^e(Z) + c_L^e(Z) \right) + \tfrac{1}{4}|r_Z|^2 \left( c_R^f(Z)^2 + c_L^f(Z)^2 \right)\left( c_R^e(Z)^2 + c_L^e(Z)^2 \right)$$
$$= 2 Q_f \operatorname{Re}(r_Z) c_V^f(Z) c_V^e(Z) + |r_Z|^2 \left( c_V^f(Z)^2 + c_A^f(Z)^2 \right)\left( c_V^e(Z)^2 + c_A^e(Z)^2 \right) \tag{7.35}$$

$$A_0(P) = \tfrac{1}{2} Q_f \operatorname{Re}(r_M) \left( c_R^f(P) + c_L^f(P) \right)\left( c_R^e(P) + c_L^e(P) \right) + \tfrac{1}{4}|r_M|^2 \left( c_R^f(P)^2 + c_L^f(P)^2 \right)\left( c_R^e(P)^2 + c_L^e(P)^2 \right)$$
$$= 2 Q_f \operatorname{Re}(r_M) c_V^f(P) c_V^e(P) + |r_M|^2 \left( c_V^f(P)^2 + c_A^f(P)^2 \right)\left( c_V^e(P)^2 + c_A^e(P)^2 \right) \tag{7.36}$$

$$A_0(Z \Leftrightarrow P) = \tfrac{1}{2} \operatorname{Re}(r_Z) \operatorname{Re}(r_M) \left( c_R^f(Z) c_R^f(P) + c_L^f(Z) c_L^f(P) \right)\left( c_R^e(Z) c_R^e(P) + c_L^e(Z) c_L^e(P) \right)$$
$$= 2 \operatorname{Re}(r_Z) \operatorname{Re}(r_M) \left( c_V^f(Z) c_V^f(P) + c_A^f(Z) c_A^f(P) \right)\left( c_V^e(Z) c_V^e(P) + c_A^e(Z) c_A^e(P) \right) \tag{7.37}$$

$$A_1 = A_1(Z) + A_1(P) + A_1(Z \Leftrightarrow P) \tag{7.38}$$

$$A_1(Z) = Q_f \operatorname{Re}(r_Z) \left( c_R^f(Z) - c_L^f(Z) \right)\left( c_R^e(Z) - c_L^e(Z) \right) + \tfrac{1}{2}|r_Z|^2 \left( c_R^f(Z)^2 - c_L^f(Z)^2 \right)\left( c_R^e(Z)^2 - c_L^e(Z)^2 \right)$$
$$= 4 Q_f \operatorname{Re}(r_Z) c_A^f(Z) c_A^e(Z) + 8|r_Z|^2 c_V^f(Z) c_A^f(Z) c_V^e(Z) c_A^e(Z) \tag{7.39}$$

$$A_1(P) = Q_f \operatorname{Re}(r_M) \left( c_R^f(P) - c_L^f(P) \right)\left( c_R^e(P) - c_L^e(P) \right) + \tfrac{1}{2}|r_M|^2 \left( c_R^f(P)^2 - c_L^f(P)^2 \right)\left( c_R^e(P)^2 - c_L^e(P)^2 \right)$$
$$+ 4 Q_f \operatorname{Re}(r_M) c_A^f(P) c_A^e(P) + 8|r_M|^2 c_V^f(P) c_A^f(P) c_V^e(P) c_A^e(P) \tag{7.40}$$

$$A_1(Z \Leftrightarrow P) = \operatorname{Re}(r_Z) \operatorname{Re}(r_M) \left( c_R^f(Z) c_R^f(P) - c_L^f(Z) c_L^f(P) \right)\left( c_R^e(Z) c_R^e(P) - c_L^e(Z) c_L^e(P) \right)$$
$$= 2 \operatorname{Re}(r_Z) \operatorname{Re}(r_M) \left( c_A^f(Z) c_V^f(P) + c_V^f(Z) c_A^f(P) \right)\left( c_A^e(Z) c_V^e(P) + c_V^e(Z) c_A^e(P) \right) \tag{7.41}$$

Then, for $p_f = p_i$, making use of (7.33), we can calculate the differential cross section:

$$\left. \frac{d\sigma}{d\Omega} \right|_{cm} = \frac{1}{64\pi^2 s} \frac{p_f}{p_i} |\mathfrak{M}_A + \mathfrak{M}_A + \mathfrak{M}_M|^2 = \frac{a_e'^2}{4s} \left[ A_0 (1 + \cos^2 \theta) + A_1 \cos \theta \right]. \tag{7.42}$$

Integrating the above over $d\Omega$, we obtain the total, unpolarized cross section:



$$\sigma(e\bar{e} \to f\bar{f}) = \frac{4\pi a_e^{'2} A_0}{3s} \equiv \sigma_0 A_0 \tag{7.43}$$

or alternatively, the cross section ratio:

$$\frac{\sigma(e\bar{e} \to f\bar{f})}{\sigma_0} = A_0 = Q_f^2 + A_0(Z) + A_0(P) + A_0(Z \Leftrightarrow P)$$
$$= Q_f^2 + 2Q_f \operatorname{Re}(r_Z)c_V^f(Z)c_V^e(Z) + |r_Z|^2 \left(c_V^f(Z)^2 + c_A^f(Z)^2\right)\left(c_V^e(Z)^2 + c_A^e(Z)^2\right) \tag{7.44}$$
$$+ 2Q_f \operatorname{Re}(r_M)c_V^f(P)c_V^e(P) + |r_M|^2 \left(c_V^f(P)^2 + c_A^f(P)^2\right)\left(c_V^e(P)^2 + c_A^e(P)^2\right)$$
$$+ 2\operatorname{Re}(r_Z)\operatorname{Re}(r_M)\left(c_V^f(Z)c_V^f(P) + c_A^f(Z)c_A^f(P)\right)\left(c_V^e(Z)c_V^e(P) + c_A^e(Z)c_A^e(P)\right)$$

We see in the above, that with the magnetic monopole interactions included, we find the usual electroweak term $Q_f^2 + A_0(Z)$ on the second line of (7.44), as well as a new term $A_0(P)$ for magnetic monopole interaction terms, and also a new term $A_0(Z \Leftrightarrow P)$ which is a cross term between the weak neutral current mediated by $Z^\mu$ and the magnetic monopole current mediated by $M^\mu$. By separating the terms in this way, we will be able to clearly see how each of the electromagnetic, weak neutral current, and magnetic monopole interactions enhance the cross section at various energies. For low energy, $|r_Z|, |r_M| \to 0$ (7.44) reduces to:

$$\frac{\sigma(e\bar{e} \to f\bar{f})}{\sigma_0} = A_0 = Q_f^2, \tag{7.45}$$

which is the cross section for ordinary, low energy, electromagnetic interactions. This is another way of stating the fact that we do not observe the magnetic monopole charge and its interactions at low energies.

Now we turn to examine the cross section enhancements due to the weak $Z^\mu$ and the $M^\mu$ of the magnetic monopole interaction, with a focus on $Z^\mu \Leftrightarrow M^\mu$ interference effects.

## 8. Cross-Section Enhancements at $M_Z$ and $M_M$

Now, let us calculate the cross sections for $e\bar{e} \to f\bar{f}$ decay to all four flavors of Fermion. We include the neutrino even though it does not have a charge, for reasons that will shortly become apparent.

We first use (7.44) together with (5.11) through (5.14) to write (the factor of 3 in (8.3) and (8.4) is for three quark colors):

$$\frac{\sigma(e\bar{e} \to \nu\bar{\nu})}{\sigma_0} = .126|r_Z|^2, \tag{8.1}$$



$$\frac{\sigma(e\bar{e} \to \mu\bar{\mu})}{\sigma_0} = 1 - .0028\,\text{Re}(r_Z) + .0632|r_Z|^2 - 2\,\text{Re}(r_M) + 4|r_M|^2 + .5780\,\text{Re}(r_Z)\,\text{Re}(r_M),  \tag{8.2}$$

$$\frac{\sigma(e\bar{e} \to u\bar{u})}{\sigma_0} = 3 \cdot \left[\tfrac{4}{9} - .0096\,\text{Re}(r_Z) + .0723|r_Z|^2 - \tfrac{8}{9}\,\text{Re}(r_M) + \tfrac{16}{9}|r_M|^2 + .4960\,\text{Re}(r_Z)\,\text{Re}(r_M)\right], \tag{8.3}$$

$$\frac{\sigma(e\bar{e} \to d\bar{d})}{\sigma_0} = 3 \cdot \left[\tfrac{1}{9} - .0087\,\text{Re}(r_Z) + .0929|r_Z|^2 - \tfrac{2}{9}\,\text{Re}(r_M) + \tfrac{4}{9}|r_M|^2 + .3031\,\text{Re}(r_Z)\,\text{Re}(r_M)\right], \tag{8.4}$$

Let us first consider these cross sections on Z-mass shell, at $s = p^\sigma p_\sigma = M_Z^2$. We also use $M_Z = 91.1876\,GeV$ as well as the experimental value $\Gamma(Z) = 2.4952\,GeV$, see following (6.6), and $\sin^2\theta_W(M_Z) = .23120$. From (7.22):

$$r_Z = \tfrac{1}{4}\frac{g_z^2}{s - M_Z^2 + iM_Z\Gamma_Z}\frac{s}{g_e'^2} = -\tfrac{1}{4}i\frac{M_Z}{\Gamma_Z}\frac{g_z^2}{g_e'^2} = -\tfrac{1}{4}i\frac{M_Z}{\Gamma_Z}\frac{1}{\sin^2\theta_W \cos^2\theta_W} = -51.4142i. \tag{8.5}$$

and therefore:

$$\text{Re}(r_Z) = 0; \quad |r_Z|^2 = 2643.4190 \tag{8.6}$$

From (7.23), we obtain, also using $\frac{g_m'^2}{g_e'^2} = \frac{1}{4a_e'^2}$, $a_e' \cong \frac{1}{128}$, $M_M \cong 2.35\,TeV$, and $\Gamma_M = 98.64\,TeV$:

$$r_M = \tfrac{1}{4}\frac{g_m'^2}{s - M_M^2 + iM_M\Gamma_M}\frac{s}{g_e'^2} = \tfrac{1}{16}\frac{M_Z^2}{M_Z^2 - M_M^2 + iM_M\Gamma_M}\frac{1}{a_e'^2}$$
$$= \tfrac{1}{16}\frac{91^2}{91^2 - 2350^2 + i2350\cdot 98640}128^2 = \frac{8{,}479{,}744}{-5{,}514{,}219 + i231{,}804{,}000} \tag{8.7}$$

Therefore:

$$\text{Re}(r_M) = -1.5378; \quad |r_M|^2 = .0013. \tag{8.8}$$

Now we return to (8.1) through (8.4) to write:

$$\frac{\sigma(e\bar{e} \to \nu\bar{\nu})}{\sigma_0}(s = M_Z^2) = .126|r_Z|^2 = .126 \cdot 2643.4190 = 333.0708, \tag{8.9}$$



$$\frac{\sigma(e\bar{e} \to \mu\bar{\mu})}{\sigma_0}(s = M_Z^2) = 1 - .0028\,\text{Re}(r_Z) + .0632|r_Z|^2 + \left[-2\,\text{Re}(r_M) + 4|r_M|^2 + .5780\,\text{Re}(r_Z)\,\text{Re}(r_M)\right], \quad (8.10)$$
$$= 168.0641 + [3.0808] = 171.1449$$

$$\frac{\sigma(e\bar{e} \to u\bar{u})}{\sigma_0}(s = M_Z^2) = \tfrac{4}{3} - .0288\,\text{Re}(r_Z) + .2169|r_Z|^2 + \left[-\tfrac{8}{3}\,\text{Re}(r_M) + \tfrac{16}{3}|r_M|^2 + 1.488\,\text{Re}(r_Z)\,\text{Re}(r_M)\right], \quad (8.11)$$
$$= 574.6910 + [4.1077] = 578.7987$$

$$\frac{\sigma(e\bar{e} \to d\bar{d})}{\sigma_0}(s = M_Z^2) = \tfrac{1}{3} - .0261\,\text{Re}(r_Z) + .2787|r_Z|^2 + \left[-\tfrac{2}{3}\,\text{Re}(r_M) + \tfrac{4}{3}|r_M|^2 + .9093\,\text{Re}(r_Z)\,\text{Re}(r_M)\right]. \quad (8.12)$$
$$= 737.0542 + [1.0291] = 738.0833$$

In square brackets, we have segregated the additional enhancements due to the magnetic monopole interaction, over and above what is ordinarily expected from the electroweak interaction in the standard model. For the electron, the magnetic monopole interaction enhances the cross section by 3.0808/168.0641 = 1.8331%. For the up quark, the enhancement is 0.7139%. For the down it is 0.1395%. For the neutrino, of course, there is no enhancement. In fact, it is apparent that this works the other way too. That is, for $v\bar{v} \to f\bar{f}$, we see from (8.9) that *no cross-section enhancement is to be expected, because the neutrino does not carry an electric or magnetic charge*. Thus, $e\bar{e} \to \mu\bar{\mu}, q\bar{q}$ interactions will show an enhanced cross section due to magnetic charges, while $v\bar{v} \to f\bar{f}$ shows no such enhancement. As we shall see in the next section, these enhancements, if their origin were not known to be due to magnetic monopoles, might be attributable instead to some type of anomaly in the weak mixing angle.

For the moment, however, let us consider these same cross section on M-mass shell, that is, at $s = M_M^2 \sim 2.35\,TeV$. Here, we use $a_e' \cong \frac{1}{126}$ as discussed earlier. From (7.22):

$$r_Z = \tfrac{1}{4}\frac{s}{s - M_Z^2 + iM_Z\Gamma_Z}\frac{1}{\sin^2\theta_W \cos^2\theta_W} = \tfrac{1}{16}\frac{M_M^2}{M_M^2 - M_Z^2 + iM_Z\Gamma_Z}\frac{1}{\sin^2\theta_W \cos^2\theta_W}. \quad (8.13)$$
$$\tfrac{1}{16}\frac{5,522,500}{5,514,184 + i227.0632}\frac{1}{.1777} = \frac{1,942,353.6860}{5,514,184 + i227.0632}$$

Therefore:

$$\text{Re}(r_Z) = \frac{1,942,353.6860}{5,514,184} = .3522\,;\text{ and }\,|r_Z|^2 = \frac{1,942,353.6860^2}{5,514,184^2 + 227.0632^2} = .1241 \quad (8.14)$$

Similarly, from (7.23):

$$r_M = \tfrac{1}{4}\frac{s}{s - M_M^2 + iM_M\Gamma_M}\frac{g_m'^2_\tau}{g_e'^2} = \tfrac{1}{16}\frac{s}{s - M_M^2 + iM_M\Gamma_M}\frac{1}{a_e'^2} = -i\tfrac{1}{16}\frac{M_M}{\Gamma_M}\frac{1}{a_e'^2} = 23.6156i. \quad (8.15)$$



Therefore:

$$\text{Re}(r_M) = 0; \quad |r_M|^2 = 577.6942 \qquad (8.16)$$

Returning again to (8.1) to (8.4), we now obtain:

$$\frac{\sigma(e\bar{e} \to \nu\bar{\nu})}{\sigma_0}\left(s = M_M^2\right) = .126|r_Z|^2 = .0156, \qquad (8.17)$$

$$\frac{\sigma(e\bar{e} \to \mu\bar{\mu})}{\sigma_0}\left(s = M_M^2\right) = 1 - .0028\,\text{Re}(r_Z) + .0632|r_Z|^2 + \left[-2\,\text{Re}(r_M) + 4|r_M|^2 + .5780\,\text{Re}(r_Z)\,\text{Re}(r_M)\right], \qquad (8.18)$$
$$= 1.0069 + [2230.7768] = 2231.7837$$

$$\frac{\sigma(e\bar{e} \to u\bar{u})}{\sigma_0}\left(s = M_M^2\right) = \tfrac{4}{3} - .0288\,\text{Re}(r_Z) + .2169|r_Z|^2 + \left[-\tfrac{8}{3}\,\text{Re}(r_M) + \tfrac{16}{3}|r_M|^2 + 1.488\,\text{Re}(r_Z)\,\text{Re}(r_M)\right], \qquad (8.19)$$
$$= 1.3501 + [3081.0357] = 3082.3858$$

$$\frac{\sigma(e\bar{e} \to d\bar{d})}{\sigma_0}\left(s = M_M^2\right) = \tfrac{1}{3} - .0261\,\text{Re}(r_Z) + .2787|r_Z|^2 + \left[-\tfrac{2}{3}\,\text{Re}(r_M) + \tfrac{4}{3}|r_M|^2 + .9093\,\text{Re}(r_Z)\,\text{Re}(r_M)\right]. \qquad (8.20)$$
$$= .3606 + [770.2589] = 770.6195$$

It is clear that at $s = M_M^2 \sim 2.35\,TeV$, there are very sizable cross-section enhancements to be expected over and above what is expected from electroweak theory. In fact, the electroweak enhancements at this energy are virtually negligible in comparison with the magnetic monopole enhancements. So, clearly, if these massive mediators do exist at $M_M^2 \sim 2.35\,TeV$,[*] their interactions will most certainly provide a good target at for the TeV-range accelerators at Fermilab and LHC.

For the moment, however, let us turn back to the much smaller enhancements predicted on mass shell at $s = M_Z^2 = 91.1876\,GeV$ in (8.10) to (8.12), because although small, it appears that these may already have been observed in relation to the NuTeV anomaly.

## 9. A Possible Connection to the NuTeV Anomaly

Returning to (8.10) for the electrons, let us suppose – not knowing about the magnetic monopole interactions first introduced in [1] and further developed here – that we tried to attribute the entire factor of 171.1449 to the electroweak neutral current. That is, if we were unaware of the magnetic monopole-based term $-2\,\text{Re}(r_M) + 4|r_M|^2 + .5780\,\text{Re}(r_Z)\,\text{Re}(r_M)$ in

---

[*] Again, assuming that $v = v_F = 246.220$ GeV.



(8.10), and were we to observe this $\frac{\sigma(e\bar{e} \to \mu\bar{\mu})}{\sigma_0} = 171.1449$ ratio in an on-shell experiment, $s = p^{\sigma} p_{\sigma} = M_Z^2$, we would be led to conclude that:

$$\frac{\sigma(e\bar{e} \to \mu\bar{\mu})}{\sigma_0} = 1 + .0632|r_Z|^2 = 171.1449 \text{, i.e., } |r_Z|^2 = \frac{170.1449}{.0632} = 2692.6883 \tag{9.1}$$

From (8.6), we also know that:

$$|r_Z|^2 = \tfrac{1}{16}\left(\frac{M_Z}{\Gamma_Z}\right)^2 \frac{1}{\sin^4\theta_W \cos^4\theta_W} = \tfrac{1}{16}(36.5452)^2 \frac{1}{\sin^4\theta_W \cos^4\theta_W} = 83.4720 \frac{1}{\sin^4\theta_W \cos^4\theta_W}. \tag{9.2}$$

This is why we used the experimental values $M_Z = 91.1876\,GeV$ and $\Gamma(Z) = 2.4952\,GeV$ in this calculation, namely, so we could isolate the impact of this discrepancy into $\sin^2\theta_W$.

If we now contrast (9.1) with (9.2), we find that:

$$|r_Z|^2 = 2692.6883 = 83.4720 \frac{1}{\sin^4\theta_W \cos^4\theta_W} \text{, i.e., } \sin^2\theta_W \cos^2\theta_W = .1761 \tag{9.3}$$

If we then write:

$$\sin^2\theta_W(1 - \sin^2\theta_W) = \sin^2\theta_W - \sin^4\theta_W = .1761 \text{, i.e., } \sin^4\theta_W - \sin^2\theta_W + .1761 = 0 \tag{9.4}$$

we find that this is quadratic in $\sin^2\theta_W$, so that for the negative root:

$$\sin^2\theta_W = \frac{1 \pm \sqrt{1 - 4 \cdot .1761}}{2} = .2282 \tag{9.5}$$

However, this is *not* the $\sin^2\theta_W(M_Z)$ we started with. Rather, we began with PDG's $\sin^2\theta_W(M_Z) = .23120$, so on mass shell at $M_Z$, this is a difference of $\Delta\sin^2\theta_W(M_Z) = -.0030$. The neutrino, of course, carries no electric or magnetic charge at all and so yields no cross section enhancement based on magnetic monopoles, see (8.9). Therefore, if we did not know that $-2\text{Re}(r_M) + 4|r_M|^2 + .5780\,\text{Re}(r_Z)\text{Re}(r_M)$ was responsible in (8.10) for adding a 1.8331% enhancement factor to the $e\bar{e} \to \mu\bar{\mu}$ cross section beyond the weak neutral current, we would conclude that the value of $\sin^2\theta_W(M_Z)$ for electron interactions is anomalously-less than that for neutrino interactions by .0030. In other words, $\Delta\sin^2\theta_W(M_Z) = -.0030$ is another way of expressing the cross section enhancement due to magnetic monopoles, but this is really not a change in the weak mixing angle but an effect due to $Z^\mu \Leftrightarrow M^\mu$ interference. But, as possible good fortune may have it, this is right around the magnitude of the NuTeV anomaly. Not only that, but the direction of this anomaly is also correct, that is, the NuTeV anomaly suggests a



higher $\sin^2 \theta_W$ for neutrino interactions over electron and quark interactions. In this light, *it appears that the observation of the NuTeV anomaly in the weak mixing angle, is perhaps the first experimental evidence of the existence of magnetic monopoles.*

Let us perform a similar calculation for the up and down quarks, to see what sort of anomaly in the mixing angle may be expected for these fermions. For the up quarks we employ (8.11), but similarly to (9.1), if one were to attribute the cross section enhancement fully to $Z^\mu$ interactions, we would write:

$$\frac{\sigma(e\bar{e} \to u\bar{u})}{\sigma_0} = \tfrac{4}{3} + .2169|r_Z|^2 = 578.7987 \text{ , i.e., } |r_Z|^2 = \frac{578.7987 - \tfrac{4}{3}}{.2169} = 2662.3576 \qquad (9.6)$$

Equation (9.2) applies intact here, without change, so combining (9.6) and (9.2) we obtain:

$$|r_Z|^2 = 83.4720 \frac{1}{\sin^4 \theta_W \cos^4 \theta_W} = 2662.3576 \text{, i.e., } \sin^2 \theta_W \cos^2 \theta_W = .1771 \qquad (9.7)$$

Now, (9.5) becomes:

$$\sin^2 \theta_W = \frac{1 \pm \sqrt{1 - 4 \cdot .1771}}{2} = .2300 \qquad (9.8)$$

which yields a difference of $\Delta \sin^2 \theta_W(M_Z) = -.0012$ from the original $\sin^2 \theta_W(M_Z) = .23120$.

For the down quark, were the enhancement attributed solely to $Z^\mu$, (8.12) gives us:

$$\frac{\sigma(e\bar{e} \to d\bar{d})}{\sigma_0} = \tfrac{1}{3} + .2787|r_Z|^2 = 738.0833 \text{, i.e., } |r_Z|^2 = \frac{738.0833 - \tfrac{1}{3}}{.2787} = 2647.1115 \qquad (9.9)$$

Therefore:

$$|r_Z|^2 = 83.4720 \frac{1}{\sin^4 \theta_W \cos^4 \theta_W} = 2647.1115 \text{, i.e., } \sin^2 \theta_W \cos^2 \theta_W = .1776 \qquad (9.10)$$

which leads to:

$$\sin^2 \theta_W = \frac{1 \pm \sqrt{1 - 4 \cdot .1776}}{2} = .2309 \qquad (9.11)$$

for a much smaller anomaly, $\Delta \sin^2 \theta_W(M_Z) = -.0003$ for the down quark. In general, the anomaly drives the weak mixing angle downward roughly with the square of the charge of the Fermion in question.

Now, it is important to emphasize that while we have employed $\sin^2 \theta_W(M_Z) = .23120$ here, what is important is not the exact magnitude of $\sin^2 \theta_W(M_Z)$, but rather, the size and



direction of the "apparent reduction" in this angle. We call this an "apparent reduction," because the weak mixing angle is not actually reduced. Rather, the NuTeV experiments may be picking up the first detectable, low energy hints of magnetic monopole interactions, which may be slightly but definitively enlarging the cross sections over what one would expect from electroweak interactions in isolation, again, see (8.10) to (8.12). While some speculations suggest that the NuTeV anomaly is due to some peculiar property of the neutrino, the result suggests that $\sin^2 \theta_W$ is given its firmest footing if it is based on $\nu\bar{\nu} \rightarrow$ decays, rather than on $e\bar{e} \rightarrow$ or any other decays, because the neutrinos – lacking a magnetic charge – do not introduce the complication of having a magnetic charge and therefore giving rise to magnetic monopole interference. That is, *it is not the neutrinos which cause the anomaly, but all the other fermions, because they all carry a magnetic charge while the neutrino does not*. We may summarize this "apparent reduction" on shell at $M_Z$, irrespective of the exact magnitude of $\sin^2 \theta_W$ itself, by saying that the magnetic monopole interactions developed here, based on [1], predict that:

$$\Delta \sin^2 \theta_W (\nu\bar{\nu} \rightarrow \mu\bar{\mu})(M_Z) - \Delta \sin^2 \theta_W (e\bar{e} \rightarrow \mu\bar{\mu})(M_Z) = +.0030 \tag{9.12}$$

$$\Delta \sin^2 \theta_W (\nu\bar{\nu} \rightarrow u\bar{u})(M_Z) - \Delta \sin^2 \theta_W (e\bar{e} \rightarrow u\bar{u})(M_Z) = +.0012 \tag{9.13}$$

$$\Delta \sin^2 \theta_W (\nu\bar{\nu} \rightarrow d\bar{d})(M_Z) - \Delta \sin^2 \theta_W (e\bar{e} \rightarrow d\bar{d})(M_Z) = +.0003 \tag{9.14}$$

$$\sin^2 \theta_W (\nu\bar{\nu} \rightarrow \mu\bar{\mu}) = \sin^2 \theta_W (\nu\bar{\nu} \rightarrow u\bar{u}) = \sin^2 \theta_W (\nu\bar{\nu} \rightarrow d\bar{d}) \tag{9.15}$$

## 10. Conclusion

The results here, particularly as regards the NuTeV anomaly, are very preliminary, and need to be explored more carefully before it can be known whether they provide a full or partial or any explanation of the NuTeV anomaly. Right now, it is noteworthy that the result (9.12) appears to be of a plausible magnitude and sign. Additionally, it has been observed that the anomaly for quarks is smaller than that for leptons, and the spread of .0018 between (9.13) and (9.12) and .0027 between (9.14) and (9.12) appears to also be within the "final" experimental errors in [6] at page 6. If these features of the "apparent reduction" in $\sin^2 \theta_W (M_Z)$ are borne out by detailed experimental study on mass shell, then an important next step will be to see what happens off mass shell, away from the $M_Z$ poles. As higher energies are observed in the TeV range, this "anomaly" will become much more pronounced. As shown (8.17) to (8.20), especially as we approach $s = M_Z^2$, the magnetic monopole interactions will become dominant and the $Z^\mu$ will make only a miniscule contribution. But, there should also be off-shell effects even below the $M_Z$ pole, i.e., at $s < M_Z^2$, and it would be interesting to see if these results can be fitted to the off-pole data as well.*

It must be noted that several hypotheses were used to reach these particular numeric results. First and foremost, we have assumed that the vev for duality is the same as the Fermi vev for electroweak theory, i.e., that $v = v_F = 246.220$ GeV, see [1], section 8. In addition, in section 4 here, certain hypotheses were made about the nature of the chiral symmetries for interactions of magnetic charges, see (4.4), (4.5), (4.9). At bottom, the exact magnitude of the cross section enhancement at given $s \equiv p^\tau p_\tau$ due to magnetic monopole interactions – and therefore the magnitude of the "anomaly" in $\sin^2 \theta_W$ – depends on six parameters: the four vertex factors $c_V^f(P)$, $c_V^e(P)$, $c_A^f(P)$, $c_A^e(P)$, the mass $M_M$, and the width $\Gamma_M$, see (7.44) and (7.23). The latter two parameters, however, depend on $v$ as well as the charge $g_m'$ and the vertex factors, see (6.7) through (6.10). So, in fact, the six theoretical parameters which ultimately determine the anomaly are $c_V^f(P)$, $c_V^e(P)$, $c_A^f(P)$, $c_A^e(P)$, $v$ and $g_m'$.

---

* If the fermions do carry magnetic charge as suggested here, one would certainly be prudent to explore the possible impact on magnetic moments as well, such as anomalies recently observed for the muon.

The magnitude of the magnetic monopole charge chiral asymmetry, if any, can be deduced via all of the terms (7.35) to (7.37) and (7.39) to (7.41), but left versus right-handedness is determined by the signs of $c_A^f(P)$, $c_A^e(P)$, and so must be picked up from the cross terms $A_0(Z \Leftrightarrow P)$ of (7.37) and $A_1(Z \Leftrightarrow P)$ of (7.41), since these cross terms are *not* invariant under the sign reversal $c_A(P) \to -c_A(P)$. Conversely, (7.35), (7.36), (7.39) and (7.40) cannot be used to determined handedness because these are invariant under $c_A(P) \to -c_A(P)$ since $c_A^f(P)$, $c_A^e(P)$ appear only as $c_A^f(P)^2$ and $c_A^e(P)^2$. Therefore, experimental data for the NuTeV anomaly may be used to determine whether $c_V^f(P)$, $c_V^e(P)$, $c_A^f(P)$, $c_A^e(P)$, $v$ and $g_m'$ and the hypotheses underlying their selection are correct, or whether these need to be fine tuned to achieve a more-precise match with the experimental data.

Also to be considered is the possibility that magnetic monopole interactions do indeed contribute to the NuTeV anomaly, but only in part. That is, in section 7, we based the cross section calculations on the amplitude $\mathfrak{M}_A + \mathfrak{M}_Z + \mathfrak{M}_M$. More generally, the amplitude should be taken to be $\mathfrak{M}_A + \mathfrak{M}_Z + \mathfrak{M}_M + \ldots$, where other vector mediators $X^\mu$ might also come under scrutiny.[*] Note that an amplitude $\mathfrak{M}_X$ for any new hypothesized vector boson $X^\mu$ should still depend on $c_V^f(X)$, $c_V^e(X)$, $c_A^f(X)$, $c_A^e(X)$, $M_X$, and $\Gamma_X$, and via the latter two, on a vev $v_X$ and charge $g_X'$. Thus, (7.34) through (7.42) and (7.44) provide the template to more generally consider the $\sin^2 \theta_W$ anomaly caused by one or more vector bosons with $M_X > M_Z$. Various scenarios in addition to, or even in place of, magnetic monopoles can therefore be considered and adjusted until a tight fit is achieved with the experimental data.

What is most exciting, however, is the prospect that the NuTeV anomaly might in fact be the first experimental evidence – more than 130 years after the seminal work of James Clerk Maxwell – of the existence of magnetic monopoles. Most surprising, perhaps, is the possibility that magnetic monopoles may not be separate fermions, but rather charges carried by the known Fermions which are only observed at high energies, and which are closely related to chiral symmetry. If these results can be confirmed, solutions to several diverse physics questions may start to fall into place.

---

[*] For example, the author has pointed out in [1], footnote on page 34, that the weak and strong interactions also have analogous massive vector bosons associated with "chromo-magnetic" interactions at about 1.2 to 1.3 TeV and 436 TeV respectively. Certainly, these might enhance the electroweak cross section under proper conditions, but of course, all fermions interact weakly and only quarks interact strongly so the vertex factors would have a different character than those elaborated here.

# References

[1] Yablon, J. R., *Magnetic Monopoles and Duality Symmetry Breaking in Maxwell's Electrodynamics*, hep-ph/0508257 (August 24, 2005).

[2] Volovik, G. E., *The Universe in a Helium Droplet*, Clarendon Press – Oxford (2003).

[3] S. Eidelman et al., Phys. Lett. B 592, 1 (2004), *Physical Constants*, http://pdg.lbl.gov/2004/reviews/consrpp.pdf.

[4] Halzen, F., and Martin A. D., *Quarks and Leptons: An Introductory Course in Modern Particle Physics*, J. Wiley & Sons (1984).

[5] S. Eidelman et al., Phys. Lett. B 592, 1(2004), *Gauge and Higgs Bosons*, http://pdg.lbl.gov/2004/tables/gxxx.pdf.

[6] Shaevitz, M., *"The NuTeV Anomaly"*, http://www-e815.fnal.gov/wma/shaevitz-stockholm-symposium-aug04.pdf, Stockholm Symposium on Neutrino Physics, August 2004.